\documentclass[aps,pra,twocolumn,groupedaddress,showpacs,10pt]{revtex4-1}
\usepackage{hyperref}
\usepackage{bm}
\usepackage{graphicx}

\begin{document}


\title{Reconstruction of Markovian dynamics from untimed data}


\author{Jochen Rau}
\email[]{jochen.rau@q-info.org}
\homepage[]{www.q-info.org}
\affiliation{Institute for Theoretical Physics,
University of Ulm,
Albert-Einstein-Allee 11,
89069 Ulm, 
Germany}
\affiliation{Department of Engineering,
RheinMain University of Applied Sciences,
Am Br\"uckweg 26,
65428 R\"usselsheim, 
Germany}


\date{\today}

\begin{abstract}
I develop a theoretical framework for inferring nonequilibrium equations of motion from incomplete experimental data.
I focus on genuinely irreversible, Markovian processes, for which the incomplete data are given in the form of snapshots of the macrostate at different instances of the evolution, yet without any information about the timing of these snapshots.
A reconstruction of the equation of motion must therefore be preceded by a reconstruction of time.
\end{abstract}


\maketitle




\section{\label{intro}Introduction}

Markovian processes, both reversible and irreversible, are ubiquitous.
Microscopic processes, described by the Schr\"odinger equation, are obviously Markovian;
and so are many macroscopic processes that can be described by, say, a rate, master, or Boltzmann equation.
All these processes share the common feature that they are local in time:
The state of the system at any given time fully determines its future evolution, regardless of the system's prior history;
the system exhibits no memory.
This renders Markovian processes reproducible, in the sense that preparation of the same initial state --- no matter how --- always entails the same subsequent evolution.
The ubiquity of Markovian processes is linked to the existence of disparate time scales in many systems.
The macroscopic observables whose dynamics one wishes to describe typically coincide with the slow observables, and hence they evolve on longer time scales;
whereas any memory, which is due to interaction with the other, faster degrees of freedom, fades away on a much shorter time scale.
There are situations in which this separation of time scales breaks down and memory effects do play a role \cite{PhysRev.124.983};
but in the present paper, I focus on Markovian dynamics.

Much effort has been devoted to developing theoretical frameworks that allow one to {derive} Markovian transport equations from the underlying microscopic dynamics;
and in turn, to deduce from these transport equations testable predictions for macroscopic experiments \cite{Nakajima01121958,1.1731409,Mori01031965,PhysRev.144.151,rau:physrep,balian:physrep,PhysRevE.56.6620,PhysRevE.62.4720}.
In contrast to this deductive approach, I start here from the opposite end:
I ask how one can {infer} a Markovian transport equation \textit{from experimental data}.
In particular, I consider situations where the experimental data come \textit{without time information}.
Such data may stem from past processes (say, in the geological or astronomical realm) that have left visible traces, albeit without time information;
or from processes that, again, leave visible traces but are so fast or delicate that they cannot be tracked with a clock.
Under such circumstances, before inferring the pertinent Markovian equation of motion, one must first reconstruct ``time.''
The purpose of the present paper is to show that this is possible (up to an additive and multiplicative constant);
and to furnish the necessary tools for doing so.

The reconstruction of time and of the equation of motion presupposes rather detailed knowledge about the generic structure of Markovian dynamics.
In particular, Markovian dynamics may be viewed geometrically as a flow on the manifold of macrostates.
This manifold is endowed with a rich geometric structure, and there are certain consistency conditions that any Markovian flow must satisfy.
These conditions, in conjunction with the experimental data, will turn out to constrain the form of the equation of motion just enough so that the reconstruction succeeds.
Mirroring the importance of these \textit{a priori} constraints, I start out with a comprehensive discussion of the geometry of macrostates (Sec. \ref{gibbs}) and of generic Markovian dynamics (Sec. \ref{generic}).
These introductory sections draw on ideas from the projection-operator \cite{Nakajima01121958,1.1731409,Mori01031965,PhysRev.144.151,rau:physrep}, geometric \cite{balian:physrep}, and two-generator  \cite{PhysRevE.56.6620,PhysRevE.62.4720} approaches to nonequilibrium dynamics.
Then I proceed to formulate the general prescription for reconstruction (Sec. \ref{reconstruction}), which I later illustrate with a simple example (Sec. \ref{example}).
I conclude with a brief discussion and outlook on future work, Sec. \ref{discussion}.


\section{\label{gibbs}Geometry of macrostates}

\subsection{\label{manifold}Manifold and coordinates}

To describe the static or dynamic properties of a macroscopic system,
typically only a few observables are deemed \textit{relevant} -- 
for example, the system's constants of the motion (if static), slow observables (if dynamic), or
observables pertaining to some subsystem of interest.
These relevant observables, together with the unit operator, span the so-called \textit{level of description},
a subspace within the linear space of observables \cite{rau:physrep}. 
For an arbitrary \textit{microstate} $\rho$ and level of description ${\cal G}:=\mbox{span}\{I,G_a\}$, the associated \textit{macrostate} is that state 
which, while yielding the same expectation values $\{g_a\}$ as $\rho$ for all relevant observables,
comes closest to equidistribution.
Closeness to equidistribution is measured in terms of the
\textit{von Neumann entropy} 
\begin{equation}
	S[\mu]:=-\mbox{tr}(\mu\ln\mu)
	;
\end{equation}
so the macrostate -- denoted by  $\pi(\rho)$ -- is determined  by the maximization
\begin{equation}
	\pi(\rho):=\arg \max_{\mu\in g} S[\mu]
	,
\end{equation}
where $\mu\in g$ is short for the constraints $\langle G_a\rangle_\mu=g_a \forall a$.
It has the \textit{Gibbs form}
\begin{equation}
	\pi(\rho) 
	= 
	Z(\lambda)^{-1} \exp(-\lambda^a G_a)
	,
\label{canonical}
\end{equation}
with the \textit{partition function}  
\begin{equation}
	Z(\lambda):=\mbox{tr}\{\exp(-\lambda^a G_a)\}
\end{equation}
ensuring state normalisation,
the \textit{Lagrange parameters} $\{\lambda^a\}$ adjusted such that 
$\langle G_a\rangle_{\pi(\rho)}=g_a$,
and -- for ease of notation -- 
the Einstein convention that identical upper and lower indices are to be summed over.
The operational meaning of ``relevance'' and how it leads to the Gibbs form have been discussed in Ref. \cite{PhysRevA.90.062114}.

Let ${\cal S}$ denote the set of normalized (pure or mixed) states of a given physical system;
this set constitutes a differentiable manifold.
In this manifold, the macrostates with level of description ${\cal G}$ form a submanifold, $\pi({\cal S})$.
This submanifold has dimension
$(\dim {\cal G} -1)$,
which equals the number of relevant observables 
(provided they are linearly independent).
On $\pi({\cal S})$ there are two natural choices of \textit{coordinates:} 
the relevant expectation values $\{g_a:=\langle G_a\rangle_{\pi(\rho)}\}$, or the Lagrange parameters $\{\lambda^a\}$.
The former coordinates can be expressed in terms of the latter via 
\begin{equation}
	g_a=-{\partial_a} \ln Z 
	,
\end{equation}
where $\partial_a:=\partial/\partial \lambda^a$.
Their respective gradients $\{dg_a\}$ and $\{d\lambda^a\}$ are related by
\begin{equation}
	dg_a = - C_{ab} d\lambda^b
	\ ,\ 
	d\lambda^a = - (C^{-1})^{ab} dg_b
	,
\label{gradients}
\end{equation}
with the Jacobian  (up to a sign) given by the \textit{correlation matrix}
\begin{equation}
	C_{ab}
	:=
	{\partial_a \partial_b} \ln Z
	.
\label{correlationmatrix}
\end{equation}
Associated with the expectation value coordinates $\{g_a\}$ is a \textit{local basis} of tangent vectors $\{\partial^a:=\partial/\partial g_a\}$,
related to their one-form duals $\{dg_a\}$ via $dg_b(\partial^a)=\delta^a_b$ \cite{schutz:book};
and likewise for the Lagrange parameter coordinates, with the associated local basis $\{\partial_a\}$ satisfying $d\lambda^b(\partial_a)=\delta^b_a$.

Upon infinitesimal variation of a macrostate 
its von Neumann entropy 
changes by
\begin{equation}
	dS = dS(\partial^a) \, dg_a = \lambda^a \, dg_a
	;
\label{entropydifferential}
\end{equation}
which in turn implies that the Hessian of the entropy yields (up to a sign) the inverse of the correlation matrix,
\begin{equation}
	(C^{-1})^{ab}
	=
	- \partial^a \partial^b S
	.
\end{equation}
Expectation values $x:=\langle X\rangle_{\pi(\rho)}$ of arbitrary (not necessarily relevant) observables $X$ change by
\begin{equation}
	dx
	=
	dx (\partial^b) \, dg_b
	=
	- \langle\delta G_b;X\rangle d\lambda^b
	,
\label{gradient_x}
\end{equation}
where $\delta G_b:= G_b-g_b$,
and $\langle;\rangle$ denotes the \textit{{canonical correlation function}} 
\begin{equation}
	\langle A;B\rangle:=\int_0^1 d\nu\,\mbox{tr}[{\pi(\rho)}^\nu A^\dagger {\pi(\rho)}^{1-\nu} B]
	.
\label{canonicalcorrelation}
\end{equation}
The latter constitutes a positive definite scalar product in the space of observables.
Comparing Eq. (\ref{gradient_x}) for $x=g_a$ with Eq. (\ref{gradients}), one finds that the correlation matrix can be expressed in terms of this scalar product, 
\begin{equation}
	C_{ab}=\langle \delta G_a;\delta G_b\rangle
	.
\label{correlationmatrix2}
\end{equation}

\subsection{\label{coarse}Coarse graining and projectors}

The map $\pi:{\cal S}\to\pi({\cal S})$ constitutes a \textit{coarse graining}.
While retaining information about the relevant observables, it discards all information about the rest.
Geo\-metrically, it ``projects'' a microstate $\rho\in{\cal S}$ onto the submanifold of macrostates.
Indeed, $\pi$ exhibits  typical features of a projection operator:
it is idempotent, $\pi\circ\pi = \pi$;
successive coarse grainings with respect to smaller and smaller levels of description are equivalent to a one-step coarse graining with respect to the smallest level of description,
\begin{equation}
	{\cal G} \subset {\cal F} 
	\ \Leftrightarrow \ 
	\pi_{\cal G}\circ\pi_{\cal F} = \pi_{\cal G}
	;
\end{equation}
and it is covariant under unitary transformations,
\begin{equation}
	\pi_{U{\cal G}U^\dagger}(U\rho U^\dagger) = U \pi_{\cal G}(\rho) U^\dagger
	.
\label{covariance}
\end{equation}
In contrast to an ordinary projection operator, however, this coarse graining map need not be linear.

Mirroring the coarse graining of states, there is also a coarse graining (``super-'')operator ${\cal P}$ on the space of observables.
The two are dual to each other in the sense that for arbitrary (not necessarily relevant) observables $X$ and arbitrary microstates $\rho$ it is
\begin{equation}
	\langle {\cal P} X \rangle_\rho
	=
	\langle X \rangle_{\pi(\rho)}
	\ \ \forall\ 
	X,\rho
	.
\end{equation}
The unique superoperator which satisfies this requirement, sometimes called the \textit{Robertson} \cite{PhysRev.144.151} or \textit{Kawasaki-Gunton} \cite{kawasaki+gunton} \textit{projector}, is
\begin{equation}
	{\cal P} X := x\, I + dx (\partial^a) \,\delta G_a
	.
\end{equation}
It projects arbitrary observables onto the level of description,
the projection being orthogonal with respect to the scalar product $\langle;\rangle$.
As the latter is evaluated in the macrostate $\pi(\rho)$, and hence the notion of orthogonality varies with the macrostate, the projector, too, carries an implicit dependence on the macrostate: ${\cal P}\equiv{\cal P}[\pi(\rho)]$.
Like the coarse graining operation $\pi$ on states, the projector is idempotent, ${\cal P}^2={\cal P}$;
yet unlike $\pi$, it is always linear and hence a true projector.
Its complement ${\cal Q}:={\cal I}-{\cal P}$ (with ${\cal I}$ being the unit superoperator, ${\cal I}X=X$) is also a projector and projects an arbitrary observable onto its ``irrelevant'' component.
Table \ref{lod_examples} lists some examples for different levels of description and the associated coarse graining operations $\pi$ and ${\cal P}$.

\begin{table}
\begin{center}
\begin{tabular}{c|c|c|c}
system &
discard &
$\pi(\rho)$ &
${\cal P}X$
\\
\hline
single &
coherence &
$\sum_i P_i \rho P_i$ &
$\sum_i P_i X P_i$
\\*[3pt]
$S\times E$ &
environment &
$\rho_S \otimes \iota_E$ &
$\mbox{tr}_E(\iota_E X)\otimes I_E$
\\*[3pt] 
$A\times B$ &
correlations &
$\rho_A\otimes \rho_B$ &
$
\begin{array}[t]{c}
\mbox{tr}_B(\rho_B X)\otimes I_B \\
+ I_A\otimes \mbox{tr}_A(\rho_A X) \\
- x\,I_A\otimes I_B
\end{array}
$
\\ 
\end{tabular}
\caption{Three examples of levels of description and the associated coarse graining operations $\pi$ and ${\cal P}$.
They refer to
(i) 
a single quantum system where only classical probabilities are deemed relevant, and all information about coherence is discarded;
(ii)
a system $S$ coupled to an environment $E$ where only the system properties are deemed relevant, and
all information about the environment (including system-environment correlations) is discarded;
and 
(iii)
a bipartite system where only single-particle properties are deemed relevant, and all correlations are discarded.
The $\{P_i\}$ are projection operators (on Hilbert space) pertaining to some preferred orthonormal (``decoherence'') basis. 
$\rho_i$ denotes the reduced state of subsystem $i$.
The state $\iota:=I/\mbox{tr}I$ is the totally mixed state.
In the first two examples $\pi$ is linear, and ${\cal P}$ is state-independent.
In contrast, in the third example $\pi$ is nonlinear, and ${\cal P}$ varies with the macrostate. 
\label{lod_examples}
}
\end{center}
\end{table}

\subsection{\label{riemannian}Metric and covariant derivative}

As the canonical correlation function $\langle;\rangle$ is a positive definite scalar product,
Eq. (\ref{correlationmatrix2}) implies that the correlation matrix is both symmetric and positive definite.
Therefore, the symmetric $(0,2)$ tensor field
\begin{equation}
	C:= C_{ab}\, d\lambda^a \otimes d\lambda^b
	= (C^{-1})^{ab} dg_a \otimes dg_b
\label{metrictensor}
\end{equation}
constitutes a Riemannian metric on the manifold of macrostates.
This particular metric is known as the \textit{Bogoliubov} or \textit{Kubo-Mori metric}
\cite{10.1063/1.530611,bengtsson:book}.
Up to a sign, it relates the local basis in $g$-coordinates to the one-form duals in $\lambda$-coordinates and vice versa,
\begin{equation}
	C(\partial^a) = - d\lambda^a
	\quad
	,
	\quad
	C(\partial_a) = - dg_a
	.
\label{metricandduals}
\end{equation}

The Bogoliubov-Kubo-Mori (BKM) metric has an operational meaning.
It quantifies the statistical distinguishability of nearby macrostates, in the following sense.
Two states $\rho$ and $\mu$ can be distinguished statistically if measurements on a finite sample, taken from an i.i.d. source of one state, say, $\rho$, are highly unlikely to erroneously indicate the other state, $\mu$.
The pertinent error probability is
\begin{eqnarray}
	\lefteqn{
	\mbox{prob}_{1-\epsilon}(\mu|N,\rho)
	}
	&&
	\nonumber \\
	&:=&
	\inf_{\Gamma} \left.\left\{\mbox{prob}(\Gamma|\rho^{\otimes N})\right|\mbox{prob}(\Gamma|\mu^{\otimes N})\geq 1-\epsilon\right\}
	,
	\quad
\label{defprob}
\end{eqnarray}
where $N$ denotes the size of the sample, and $\Gamma$ is a proxy for the measurement results, 
which asymptotically, i.e., to within an error probability $\epsilon$ ($0<\epsilon<1$) that does not depend on sample size, are compatible with the sample being in the state $\mu^{\otimes N}$.
Asymptotically, this error probability decreases exponentially with sample size,
\begin{equation}
	\mbox{prob}_{1-\epsilon}(\mu|N,\rho)\sim \exp[-N S(\mu\|\rho)]
	,
\label{quantumstein}
\end{equation}
and no longer depends on the specific value of the error parameter $\epsilon$ (``quantum Stein lemma'' 
\cite{hiai+petz,ogawa+nagaoka}).
The exponent features the \textit{relative entropy} \cite{donald:cmp,vedral:rmp},
\begin{equation}
	S(\mu\|\rho):=
	\left\{
\begin{array}{ll}
	\mbox{tr}(\mu\ln\mu - \mu\ln\rho) & : \mbox{supp}\;\mu\subseteq\mbox{supp}\;\rho \\
	+\infty & : \mbox{otherwise}
\end{array}
\right.
,
\label{relativeentropy}
\end{equation}
which thus proves to be a natural distinguishability measure.
For nearby macrostates $\pi(\rho)$ and $\pi(\rho+\delta\rho)$ connected by a distance vector $W$, this distinguishability measure is approximated to lowest (quadratic) order in the coordinate differentials by
\begin{equation}
	S(\pi(\rho)\|\pi(\rho+\delta\rho))
	\approx
	\textstyle\frac{1}{2} C(W,W)
	;
\label{relentropy_nearby}
\end{equation}
and hence indeed, up to a numerical factor, by the length (squared) of the distance vector in the BKM metric.

The above metric is also singled out by the fact that it is with respect to this metric that the ``projection'' $\pi$ onto the submanifold of Gibbs states is orthogonal.
This can be seen as follows.
Being the infinitesimal version of the relative entropy, and the latter being defined for arbitrary pairs of states (both macro and micro), 
the metric can be extended from the manifold of macrostates to full state space.
Let $V$ denote an arbitrary vector field on the full state space that connects only states with identical expectation values for the relevant observables, $dg_a (V)=0$.
Then with $dg_a=-C(\partial_a)$, it is
\begin{equation}
	C(\partial_a,V) = 0
	\ \ \forall\ 
	a
	;
\end{equation}
i.e., any such $V$ intersects $\pi({\cal S})$, which is generated by the basis vectors $\{\partial_a\}$, at a right angle.

A Riemannian metric allows one to raise and lower indices of tensor fields, i.e., to map an $(n,m)$ tensor field to an $(n+1,m-1)$ or $(n-1,m+1)$ tensor field, respectively, thereby preserving its total rank, $n+m$.
Henceforth I will not distinguish between tensor fields that differ only by raising or lowering of indices via the BKM metric;
I will denote such fields by the same symbol and characterize them only by their total rank.

Associated with the metric is a \textit{covariant derivative,} $\nabla$.
Given some vector field $V$, the covariant derivative \textit{along} $V$, $\nabla_V$, 
maps an arbitrary rank-$r$ tensor field $A$ to another rank-$r$ tensor field, $\nabla_V A$.
This map 
(i)
obeys the sum rule, 
$\nabla_V (A+B)=(\nabla_V A) + (\nabla_V B)$;
(ii)
obeys the Leibniz rule for tensor products,
$\nabla_V (A\otimes B)=(\nabla_V A)\otimes B + A\otimes (\nabla_V B)$;
(iii)
commutes with tensor contraction;
(iv)
is linear in the vector field, 
$\nabla_{fU+gV} A=f\nabla_U A + g\nabla_V A$, for arbitrary scalar functions $f,g$ and vector fields $U,V$.
For this reason there exists a rank-$(r+1)$ tensor field, denoted $\nabla A$ and called the \textit{gradient} of $A$, from which $\nabla_V A$ can be obtained via contraction with $V$;
and
(v)
when applied to a scalar field $\phi$,
the map coincides with the ordinary directional derivative, $\nabla_V\phi=d\phi(V)$.
So in this special case
the gradient coincides with the ordinary differential, $\nabla\phi=d\phi$.

That the covariant derivative stems from the BKM metric is reflected in the fact that 
the gradient of the metric tensor vanishes,
\begin{equation}
	\nabla C = 0
	.
\end{equation}
In particular, there is no torsion,
\begin{equation}
	\nabla_U V - \nabla_V U
	= 
	[U,V] 
	\ \ 
	\forall
	\ 
	U,V
	,
\end{equation}
the bracket $[U,V]$ being the Lie bracket of the two vector fields $U$ and $V$ \cite{schutz:book}.
The absence of torsion implies that the gradient of a one-form field $\alpha$ can be written as
\begin{equation}
	\nabla\alpha
	=
	\textstyle\frac{1}{2} [d\alpha + \pounds_{C^{-1}(\alpha)} C]
	,
\label{gradient_oneform}
\end{equation}
where $d$ denotes the exterior derivative and $\pounds$ the Lie derivative.
The first term inside the square bracket is antisymmetric, whereas the second term is symmetric.
Whenever $\alpha$ is itself a gradient, $\alpha=d\phi=\nabla\phi$, its exterior derivative vanishes, and hence it is
\begin{equation}
	\nabla\nabla\phi
	=
	\textstyle\frac{1}{2} \pounds_{C^{-1}(\nabla\phi)} C
	.
\end{equation}

One scalar function that will play a special role in my subsequent argument is the modified entropy \cite{PhysRevA.84.012101}
\begin{equation}
	S_\sigma[\mu]
	:=
	S[\sigma] - S(\mu\|\sigma)
\label{modifiedentropy}
\end{equation}
with \textit{reference macrostate} $\sigma\in\pi({\cal S})$.
This modified entropy characterizes the closeness of $\mu$ to the reference macrostate;
it reduces to the ordinary von Neumann entropy when the reference macrostate equals the totally mixed state.
Upon infinitesimal variation of a macrostate (at fixed reference macrostate) the modified entropy changes in a manner similar to the ordinary entropy, Eq. (\ref{entropydifferential}), only with $\lambda^a$ replaced by $(\lambda^a-\lambda^a_\sigma)$,
\begin{equation}
	dS_\sigma = (\lambda^a-\lambda^a_\sigma) dg_a
	,	
\end{equation}
where the $\{\lambda^a_\sigma\}$ pertain to the reference macrostate $\sigma$.
Taking the gradient of this differential yields, with the help of Eqs. (\ref{gradients}) and (\ref{metrictensor}),
\begin{equation}
	\nabla\nabla S_\sigma = -C + (\lambda^a-\lambda^a_\sigma) \nabla\nabla g_a
	.
\label{Hessianandmetric}
\end{equation}
So when evaluated at the reference macrostate, where $\lambda^a=\lambda^a_\sigma$, this covariant Hessian equals (up to a sign) the metric tensor,
\begin{equation}
	\nabla\nabla S_\sigma = -C
	\ 
	\mbox{at}\ \sigma
	.
\label{metricasgradient}
\end{equation}

When macrostates are constrained to the hyperplane $\Sigma:=\{\mu\in\pi({\cal S})|\langle\ln\sigma\rangle_\mu=\langle\ln\sigma\rangle_\sigma\}$,
modified and ordinary entropies coincide,
$S_\sigma|_\Sigma=S|_\Sigma$.
Then Eq. (\ref{metricasgradient}) carries over to the ordinary entropy,
\begin{equation}
	\nabla\nabla S|_\Sigma
	=
	- C|_\Sigma
	\ 
	\mbox{at}
	\ 
	\sigma
	.
\label{metricasgradient_constrained}
\end{equation}
As an example, $\sigma$ might be a canonical equilibrium state and hence $\ln\sigma$ proportional to the Hamiltonian.
Then $\Sigma$ constitutes a hyperplane of constant energy.
It comprises all macrostates that have the same energy as $\sigma$, including $\sigma$ itself.
Whenever energy is conserved, evolution of the macrostate is constrained to such a hyperplane;
so energy-conserving flows have $C|_\Sigma$ as their relevant metric.
At the equilibrium state $\sigma$ this constrained metric is given by the (negative) Hessian of ordinary entropy.

 
\section{Generic Markovian dynamics}
\label{generic}

\subsection{Disparate time scales}

In this paper I focus on the dynamics of an isolated quantum system with time-independent Hamiltonian $H$.
(In principle, the description of an open system can be incorporated into this framework by enlarging it to include its environment.)
On the microscopic level the dynamics of such a system is governed by the \textit{Liouville-von Neumann equation}
\begin{equation}
	\dot{\rho}(t)=-i{\cal L}\rho(t)
	,
\label{Liouville_vonNeumann}
\end{equation}
where $\rho$ denotes the system's microstate, and the {\textit{Liouvillian}} ${\cal L}:=[H,\cdot]$ is a shorthand for the commutator with $H$.  
For simplicity, I set $\hbar=1$.

On the macroscopic level one seeks to describe the dynamics of only certain selected expectation values $g_a(t):=\langle G_a\rangle_{\rho(t)}$ deemed ``relevant''.
Provided that initially, at $t=0$, these relevant expectation values suffice to determine the system's microstate, i.e., $\rho(0)$ carries no information other than about $\{g_a(0)\}$ and hence has the Gibbs form
\begin{equation}
	\rho(0)
	\propto
	\exp(-\lambda^a(0) G_a)
	,
\end{equation}
their dynamics at $t\geq 0$ is governed by the \textit{Robertson equation} \cite{PhysRev.144.151,rau:physrep}
\begin{equation}
	\dot{g}_a(t) = \dot{g}_a^{(l)}(t) + \dot{g}_a^{(m)}(t)
\label{robertson}
\end{equation}
with the \textit{local term}
\begin{equation}
	\dot{g}_a^{(l)}(t)
	=
	\langle i{\cal L} G_a\rangle_{\pi(\rho(t))}
\end{equation}	
and the \textit{memory term}
\begin{equation}
	\dot{g}_a^{(m)}(t)
	=
	- \int_0^t dt'\,\langle
	{\cal L}{\cal Q}(t'){\cal T}(t',t){\cal Q}(t){\cal L} G_a
	\rangle_{\pi(\rho(t'))}
	.
\label{memoryterm}
\end{equation}
Here $\pi(\rho(t))$ is the {macrostate} at time $t$ as defined in Sec. \ref{manifold}.
The objects ${\cal Q}$ and ${\cal T}$ are, like the Liouvillian, superoperators acting on the space of observables.
The former is a projector, ${\cal Q}^2={\cal Q}$, which projects any observable onto its ``irrelevant'' component;
it is the complement of the Robertson projector defined in Sec. \ref{coarse}.
Like the Robertson projector, it may vary with the macrostate and thus may carry an implicit time dependence.
The superoperator ${\cal T}$ effects the time evolution of irrelevant degrees of freedom,
\begin{equation}
	(\partial/\partial t') {\cal T}(t',t)
	=
	-i {\cal Q}(t') {\cal L} {\cal Q}(t') {\cal T}(t',t)
	,
\end{equation}
with initial condition ${\cal T}(t,t)={\cal I}$.

All terms on the right-hand side of the Robertson equation depend on the macrostate and hence on relevant expectation values only;
so the system of equations of motion for the $\{g_a(t)\}$ is indeed closed.
Irrelevant degrees of freedom have been eliminated completely from the description of the macroscopic dynamics.
The price to pay for this elimination is that in contrast to the microscopic Liouville-von Neumann equation, the Robertson equation can be both nonlocal in time and nonlinear. 
The former means that the change of relevant expectation values at any given time may depend not just on their current values but on their entire history since $t=0$,
i.e., that the macroscopic dynamics has a {``memory''}.
The latter -- nonlinearity -- may arise whenever the coarse graining operation $\pi$ is not linear.
A simple example for such a nonlinear coarse graining was given in Table \ref{lod_examples}.
Indeed, many well-known transport equations such as the Boltzmann or Navier-Stokes equations, which can be derived within the above framework or its classical counterpart, are nonlinear.

The Robertson equation becomes \textit{Markovian}, i.e., local in time, if and only if the physical system exhibits a clear separation of time scales, and it is the slow degrees of freedom which are chosen as the relevant ones.
In this case the ``memory time'' $\tau_m$
-- the time scale on which the integrand in Eq. (\ref{memoryterm}) falls off towards the past --
will be short compared to the time scale $\tau_r$ on which the relevant expectation values evolve.
One may then replace  
\begin{equation}
	\pi(\rho(t')) \to \pi(\rho(t))
	\ ,\ 
	{\cal Q}(t') \to {\cal Q}(t)
\end{equation}
in both the Robertson equation and the differential equation for ${\cal T}$ (``Markovian approximation'').
Moreover, provided there is a genuine gap between the two time scales, 
in the sense that there exists an intermediate scale $T$ such that
$\tau_m\ll T \ll \tau_r$,
and with the substitution $(t-t')\to\tau$
one may expand the integration range for $\tau$ from $[0,t]$ to $[0,T]$
and replace
\begin{equation}
	\int_0^t dt'\, {\cal T}(t',t)
	\ \to\ 
	{\cal I}^{(+)}
	:=
	\int_0^T d\tau\, \exp( i\tau{\cal Q}{\cal L}{\cal Q})
	,
\label{def_Iplus}
\end{equation}
where for simplicity I omitted the dependence on $t$.

Geometrically, the collected relevant expectation values at any given time can be represented as a point, and their time evolution as a curve, in the manifold of macrostates.
The curve results from projecting the trajectory of the microstate in full state space onto the lower-dimensional submanifold of macrostates.
In case the Robertson equation is Markovian (which I shall assume from now on), its general solution defines in the manifold of macrostates a congruence of curves.
This congruence in turn gives rise to a vector field on the manifold of macrostates,
\begin{equation}
	V := \dot{g}_a \partial^a
	.
\end{equation}
The correspondence being one-to-one, the Markovian dynamics may be characterized completely  by the vector field $V$.
In line with the split in Eq. (\ref{robertson}), the vector field can be broken down into contributions from the local term and the memory term, $V=V^{(l)}+V^{(m)}$.

\subsection{Effective non-dissipative dynamics}
\label{effectivedynamics}

The effective non-dissipative dynamics of the relevant expectation values is described by the local term and the antisymmetric part of the memory term.
As for the local term, 
one can use the general formula
\begin{equation}
	[X,\rho]
	=
	\int_0^1 d\nu \, \rho^{\nu} [X,\ln\rho] \rho^{1-\nu}
\end{equation}
and define on the manifold of macrostates the antisymmetric $(2,0)$ tensor field 
\begin{equation}
	K
	:=
	\langle \textstyle\frac{1}{i} [G_a,G_b] ; (H-\langle H\rangle_{\pi(\rho)}) \rangle_{\pi(\rho)} \partial^b \otimes \partial^a
\label{tensorK}
\end{equation}
to cast it into the form
\begin{equation}
	V^{(l)}
	=
	K (dS,\cdot)
	.
\end{equation}
As an immediate consequence of the antisymmetry of $K$, the local term preserves entropy,
\begin{equation}
	K(dS,dS) = 0
	,
\label{KpreservesEntropy}
\end{equation}
and is thus indeed non-dissipative.

As for the memory term,
using the relation
\begin{equation}
	\langle {\cal L} X \rangle_{\pi(\rho)}
	=
	- \lambda^b \langle {\cal L} G_b ; X \rangle_{\pi(\rho)}
\label{firsttermascorrelation}
\end{equation}
for arbitrary (not necessarily relevant) $X$,
the hermiticity of ${\cal Q}$ with respect to the canonical correlation function,
as well as the fact that the memory term must be real,
and defining the $(2,0)$ tensor field
\begin{equation}
	M:=
	\langle
	{\cal I}^{(+)} {\cal Q}{\cal L}G_a ; {\cal Q}{\cal L} G_b
	\rangle_{\pi(\rho)}
	\partial^b \otimes \partial^a
	,
\end{equation}
it can be cast into the form
\begin{equation}
	V^{(m)}
	=
	M(dS,\cdot)
	.
\end{equation}
The tensor $M$ comprises an 
antisymmetric and a symmetric part
whose respective components are given by
\begin{equation}
	M_{ab}^{(\pm)}
	:=
	\textstyle\frac{1}{2}
	(M_{ab} \pm M_{ba})
	.	
\end{equation}
The antisymmetric part conserves entropy,
\begin{equation}
	M^{(-)}(dS,dS)
	=
	0
	,
\label{M-preservesEntropy}
\end{equation}
and thus co-determines, together with the local term, the non-dissipative dynamics.
Often neglected, 
this antisymmetric part of the memory term may contain interesting physics such as geometric  or other effective forces \cite{PhysRevE.56.R1295}, and it may play an important role in Hamiltonian renormalization \cite{PhysRevE.55.5147}.
The most general Markovian dynamics is then described by the tensor
\begin{equation}
	T := (K+M^{(-)}) + M^{(+)}
	,
\end{equation}
where the first two, antisymmetric terms drive the non-dissipative dynamics, whereas the last, symmetric term drives the dissipative dynamics.
With this tensor the complete equation of motion acquires the compact form
\begin{equation}
	V=T(dS,\cdot)
	.
\label{generaldynamics}
\end{equation}

It is possible to describe the non-dissipative dynamics by an effective Hamiltonian $H_{\rm eff}$ if and only if
\begin{equation}
	(K + M^{(-)})(dS,\cdot) = K_{\rm eff}(dS,\cdot)
	,
\label{EffectiveHamiltonian}
\end{equation}
where $K_{\rm eff}$ is defined as in Eq. (\ref{tensorK}) but with $H$ replaced by $H_{\rm eff}$.
In this case the non-dissipative time evolution of macrostates is unitary.
Since unitary transformations leave relative entropies invariant and hence, thanks to Eq. (\ref{relentropy_nearby}), also the BKM metric,
it is
\begin{equation}
	\pounds_{K_{\rm eff} (dS,\cdot)} C = 0
	.
\label{unitaritycondition}
\end{equation}
Provided $H_{\rm eff}$ is itself a relevant observable,
the non-dissipative dynamics can be written in the alternative form
\begin{equation}
	K_{\rm eff} (dS,\cdot)
	=
	L (dU,\cdot)
	,
\end{equation}
where $U:=\langle H_{\rm eff}\rangle_{\pi(\rho)}$ is the (effective) internal energy, and $L$ denotes another antisymmetric $(2,0)$ tensor field,
\begin{equation}
	L := \langle \textstyle\frac{1}{i}[G_a,G_b] \rangle_{\pi(\rho)} 
	\partial^b \otimes \partial^a
	.
\label{poisson}
\end{equation}
In this formulation the conservation of entropy is reflected in the tensor property
\begin{equation}
	L(\cdot,dS)=0
	.
\end{equation}
In principle, the effective Hamilton operator may be influenced by the macrostate, $H_{\rm eff}\equiv H_{\rm eff}[\pi(\rho)]$, and thus, through the latter, depend on time.
If, however, such an explicit time dependence is absent or may be neglected, 
the antisymmetry of $L$ ensures the conservation of internal energy,
\begin{equation}
	L (dU,dU)
	=
	0
	.
\label{energyconservation}
\end{equation}

In case the relevant observables form a Lie algebra,
$(1/i)[{\cal G},{\cal G}]\subset{\cal G}$, the manifold of macrostates and their non-dissipative dynamics exhibit further structure.
Such a Lie algebra property holds for many important choices of the level of description:
for instance, when the relevant observables comprise
(i) 
all constants of the motion;
(ii)
all observables pertaining to one or several subsystems of a composite system;
or
(iii)
all block diagonal observables of the form $\sum_i P_i A P_i$, where $\{P_i\}$ is some set of mutually orthogonal projectors.
In all these examples the commutator (times $1/i$) of two relevant observables is again a relevant observable.
Then the bracket 
\begin{equation}
	\{f,g\}:= L(df,dg)
\end{equation}
of two functions $f$ and $g$, defined with the help of the tensor field $L$, possesses all properties of a Poisson bracket;
it satisfies antisymmetry, linearity, Leibniz rule, and the Jacobi identity.
Thus the manifold of macrostates is endowed with a \textit{Poisson structure} \cite{vaisman:book}.
According to the splitting theorem for Poisson manifolds \cite{weinstein1983} the manifold of macrostates can then be foliated into \textit{symplectic leaves},
each with constant entropy.
On every leaf one can define a symplectic two-form, i.e., a two-form which is antisymmetric, non-degenerate, and closed.
Also on every leaf, the vector field associated with the non-dissipative dynamics, $L (dU,\cdot)$, becomes a \textit{Hamiltonian vector field,} 
the pertinent Hamilton function being the internal energy $U$.
One recovers thus the familiar structure of classical Hamiltonian mechanics \cite{arnold:book}.

\subsection{Dissipation}

Dissipation is described by the symmetric part of the memory term.
This symmetric part is positive semidefinite, 
\begin{equation}
	M^{(+)}\geq 0
	,
\label{positivity}
\end{equation}
which can be understood as follows.
One assumes that on short time scales smaller than $T$ the dynamics of the irrelevant degrees of freedom is 
(i)
time translation invariant;
and in particular,
(ii)
unaffected by the (slow) variation of the macrostate.
Then it is
\begin{equation}
	\langle
	{\cal Q}{\cal L}G_a ; {\cal I}^{(+)} {\cal Q}{\cal L} G_b
	\rangle
	=
	\langle
	{\cal I}^{(-)} {\cal Q}{\cal L}G_a ; {\cal Q}{\cal L} G_b
	\rangle
	,
\end{equation}
where ${\cal I}^{(-)}$ is defined as in Eq. (\ref{def_Iplus}) but with a minus sign in the exponent.
Considering moreover that the memory term must be real, and hence that the canonical correlation function (which is a scalar product) featuring in the definition of $M$ must be symmetric, the components of the symmetric and antisymmetric parts of $M$ are given by
\begin{equation}
	M_{ab}^{(\pm)}
	=
	\textstyle\frac{1}{2}
	\langle
	({\cal I}^{(+)} \pm {\cal I}^{(-)}) {\cal Q}{\cal L}G_a ;  {\cal Q}{\cal L} G_b
	\rangle
	.
\end{equation}
For the symmetric part one can then write
(invoking once more the time translation invariance of the irrelevant dynamics on short time scales)
\begin{eqnarray}
	M_{ab}^{(+)}
	&=&
	\frac{1}{2}
	\int_{-T}^T d\tau\,
	\langle
	\exp(i \tau {\cal Q}{\cal L}{\cal Q}) {\cal Q}{\cal L}G_a ; {\cal Q}{\cal L} G_b
	\rangle
	\nonumber \\
	&=&
	\frac{1}{2T}
	\int_0^T ds
	\int_{-T}^T d\tau\,
	\nonumber \\
	&&
	\times
	\langle
	\exp[i (\tau+s){\cal Q}{\cal L}{\cal Q}] {\cal Q}{\cal L}G_a 
	; 
	\exp(i s{\cal Q}{\cal L}{\cal Q}) {\cal Q}{\cal L} G_b
	\rangle
	\nonumber \\
	&=&
	\frac{1}{2T}
	\int_0^T ds
	\int_{-T+s}^{T+s} d\tau'\,
	\nonumber \\
	&&
	\times
	\langle
	\exp(i \tau'{\cal Q}{\cal L}{\cal Q}) {\cal Q}{\cal L}G_a 
	; 
	\exp(i s{\cal Q}{\cal L}{\cal Q}) {\cal Q}{\cal L} G_b
	\rangle
	.
	\nonumber \\
	&& 
\end{eqnarray}
To the second integral only $\tau'\approx s \in [0,T]$ contribute significantly, and hence one may reduce its integration range $[-T+s,T+s] \to [0,T]$.
This finally yields
\begin{equation}
	M_{ab}^{(+)}
	=
	({1}/{2T})
	\langle
	{\cal I}^{(+)} {\cal Q}{\cal L}G_a ; {\cal I}^{(+)} {\cal Q}{\cal L} G_b
	\rangle
	.
\end{equation}
Since the canonical correlation function is a positive definite scalar product,
this component matrix, and hence $M^{(+)}$ itself, must indeed be positive semidefinite.

An immediate consequence of the above positivity is that the symmetric part of the memory term may only lead to an increase, but never to a decrease, of entropy,
\begin{equation}
	M^{(+)}(dS,dS)
	\geq
	0
	.
\end{equation}
This embodies the \textit{$H$-theorem}, originally formulated by Boltzmann for the dynamics of dilute gases \cite{boltzmann:h-theorem} but in fact valid for arbitrary Markovian processes.
It marks the gradual approach of the macrostate towards equilibrium.

One particularly simple type of dissipative dynamics is \textit{steepest descent} towards the equilibrium state.
Under this dynamics trajectories on the manifold of macrostates (or in case there are conservation laws: on the allowed submanifold) simply follow the entropy gradient \cite{rau:relaxation}.
The only contribution to this type of dynamics stems from the symmetric part of the memory term, which has the simple form
\begin{equation}
	M^{(+)} \propto C^{-1}
	.
\end{equation}
This is a rather natural ansatz:
The metric $C$ measures the  distance between two macrostates in the sense of their statistical distinguishability.
The symmetric part of the memory term (or its inverse, respectively), being symmetric and positive semidefinite, is a metric, too;
it measures the ``dynamical'' distance between two macrostates as mediated by interactions with the irrelevant degrees of freedom.
In the absence of any specific information about the dynamics of the irrelevant degrees of freedom, in particular about any preferred direction on the manifold of macrostates, these two metrics are taken to be equal, up to some multiplicative constant. 
A direct consequence of this ansatz is that the gradient of $M^{(+)}$ vanishes everywhere,
$\nabla M^{(+)} = 0$.
This is not the case for other, more general forms of Markovian dynamics.
The magnitude of this gradient can then be taken as a local, coordinate-independent measure for the deviation from steepest descent.

When the overall dynamics drives the macrostate towards an equilibrium macrostate $\sigma\in\pi({\cal S})$ 
then its logarithm, $\ln\sigma$, which is a relevant observable, must be a constant of the motion.
Hence in their approach towards $\sigma$, macrostates are constrained to the hyperplane $\Sigma$ defined in Sec. \ref{riemannian}.
Provided the equilibrium macrostate is a canonical state with the same effective Hamiltonian as that governing the non-dissipative dynamics, $\sigma\propto\exp(-\beta_\sigma H_{\rm eff})$, this hyperplane corresponds to fixed internal energy, $U=\mbox{const}$.
The latter is conserved by the non-dissipative dynamics, Eq. (\ref{energyconservation}), and so in order to be conserved overall, it must be conserved by the dissipative dynamics, too.
This conservation of energy is implemented mathematically by imposing 
\begin{equation}
	M^{(+)}(\cdot,dU)=0
	.
\end{equation}

\subsection{Near equilibrium}

Let $\sigma\in\pi({\cal S})$ be an equilibrium macrostate
and $\Sigma\subset\pi({\cal S})$ the associated hyperplane, defined in Sec. \ref{riemannian}, to which macrostates are constrained as they approach $\sigma$.
In the following, all mathematical objects ---macrostates, functions, vector and tensor fields--- and their relationships are meant to be constrained to this hyperplane, even if for simplicity I will omit the explicit notation ``$|_\Sigma$''.
To begin with, taking the covariant derivative on both sides of the general equation of motion, Eq. (\ref{generaldynamics}), and exploiting the relationship between the Hessian of the entropy and the metric tensor, Eq. (\ref{metricasgradient_constrained}), as well as $dS=0$ at equilibrium, yields an equation for $T$ at equilibrium,
\begin{equation}
	T = - \nabla V
	\ \mbox{at}\ \sigma
	;
\end{equation}
i.e., at equilibrium, the tensor $T$ which governs the Markovian dynamics can be gleaned from the observed vector field $V$ by taking (minus) the gradient.
To first-order approximation, this relationship can be extended to the vicinity of equilibrium, 
\begin{equation}
	T \approx - \nabla V
	;
\label{TnablaV}
\end{equation}
which in turn, by the general equation of motion, implies
\begin{equation}
	V \approx - (\nabla V) (dS,\cdot)
	.
\label{consistencycondition}
\end{equation}
The latter equation imposes a consistency condition on the vector field $V$ near equilibrium.
It goes beyond the trivial requirement that $V=0$ at $\sigma$.

By symmetrizing both sides of Eq. (\ref{TnablaV}) and exploiting the positivity of the symmetric part of $T$, Eq. (\ref{positivity}), one finds that the symmetric part of the gradient $\nabla V$ must be negative semidefinite.
This symmetric part is denoted by $[\nabla V]_+$ and may be called, in an analogy with fluid dynamics, the rate of strain tensor.
In conjunction with Eq. (\ref{gradient_oneform}), its negativity leads to the inequality
\begin{equation}
	\pounds_V C
	=
	2\,[\nabla V]_+
	\leq
	0
	,
\label{localconvergence}
\end{equation}
again valid in the vicinity of equilibrium.
Near equilibrium, therefore, macrostates converge (in terms of the BKM metric) not just towards equilibrium but also towards each other;
as they evolve, their mutual distances, and hence their statistical distinguishability, decrease monotonically.
This inequality may be viewed as a special case of Lindblad's theorem \cite{lindblad:monotonicity}, here applied to the linearized dynamics near equilibrium.

When the dynamics is entirely non-dissipative, $M^{(+)}=0$, the above reasoning implies $\pounds_V C=0$.
This is in keeping with the unitarity condition, Eq. (\ref{unitaritycondition}).
As the BKM metric is the infinitesimal version of relative entropy, Eq. (\ref{relentropy_nearby}), its conservation suggests that, moreover, arbitrary relative entropies are preserved.
If this (unproven) conjecture were true then the conservation of the BKM metric, $\pounds_V C=0$, would in fact mandate, rather than merely allow, that the non-dissipative dynamics is unitary \cite{molnar1,molnar2}.


\section{Reconstruction}
\label{reconstruction}

The general framework laid out above offers a systematic route from the microscopic to the macroscopic realm.
Starting from the microscopic Hamiltonian and the Liouville-von Neumann equation, Eq. (\ref{Liouville_vonNeumann}), it allows one, at least in principle, to calculate the tensor field $T$ that governs the Markovian macroscopic dynamics;
and from there, via the general equation of motion, the vector field $V$, and hence the time evolution of macroscopic observables.
Reconstruction starts from the other end: 
Having observed some macroscopic Markovian process experimentally, one aims to infer the dynamical law.
Specifically, one endeavours to reconstruct the pertinent tensor field $T$.

\begin{figure}[htbp]
\begin{center}
\includegraphics[width=4.8cm]{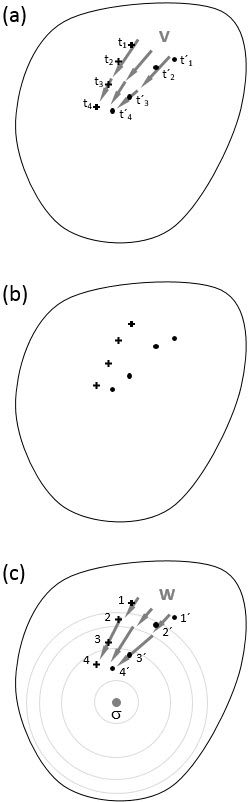}
\end{center}
\caption{Schematic illustration of the basic reconstruction idea.
The figures show a section of the hyperplane $\Sigma$ of macrostates.
(a) 
Experimental data are given in the form of timed snapshots of the macrostate at different instances of the Markovian evolution. 
Crosses and dots represent data from two different runs of the experiment with varying initial conditions;
in practice, there will be many more such runs and data points.
In combination, these data allow one to construct the vector field $V$.
(b)
When time information is removed, it is no longer possible to construct the vector field $V$.
(c)
However, as long as the dynamics is genuinely irreversible, $\dot{S}>0$, at least the temporal order of the snapshots is easily established, and one may use entropy as a proxy for time.
Here concentric circles around the equilibrium state $\sigma$ indicate hypersurfaces of constant entropy.
With this proxy one can construct a vector field $W$, related to the true $V$ by $V=\dot{S} W$. 
The entropy production rate, $\dot{S}$, can in turn be estimated from the data, up to a multiplicative constant, if one assumes the existence of an effective Hamiltonian for the non-dissipative part of the dynamics (see text).
}
\label{figure1}
\end{figure}

The generic experimental setting can be characterized as follows.
On some coarse-grained level of description, one observes a hitherto unknown process whose only known feature is that it is reproducible and hence Markovian.
The dynamical law governing this macroscopic process -- here: the tensor field $T$ -- is not known and yet to be inferred from the data.
It is impossible to infer the dynamics from a single trajectory alone;
rather, one has to study an entire family of trajectories pertaining to differently prepared test systems.
More specifically,
in order to collect sufficient data for such an inference, one must
(i)
observe multiple copies of the system of interest, 
or provided the system is large enough so that disturbances due to measurement may be neglected, observe one system undergo the same process several times;
(ii)
vary the initial macrostates of the different copies or sequential runs, respectively;
(iii)
in each run, measure the macrostate at different (ideally, closely spaced) instances of its evolution;
and 
(iv)
record the time between subsequent measurements.
In combination, these data allow one to determine the vector field $V$, as illustrated schematically in Fig. \ref{figure1}(a).

Having thus determined experimentally the vector field $V$, one proceeds to infer the macroscopic dynamical law, embodied in the tensor field $T$.
Near equilibrium, this is a straightforward exercise:
According to Eq. (\ref{TnablaV}), $T$ is but the negative gradient of $V$.
Calculating the gradient requires knowledge of the (generally non-Euclidean) geometry of the manifold of macrostates,
which depends on the level of description but is otherwise independent of the experimental data.
Further away from equilbrium, there may be corrections to Eq. (\ref{TnablaV}).
One will then end up with some system of first-order partial differential equations for the components of $T$.
The difficulty of solving these differential equations will vary depending on the system at hand.

For simplicity, I stick here to the vicinity of equilibrium and  focus instead on another issue that may complicate the reconstruction of $T$.
I consider the situation where one is given data points on the manifold of macrostates but \textit{without any information about time} (Fig. \ref{figure1}(b)).
On a large scale, 
this might be the case for, say, geological or astronomical data originating from different epochs of terrestrial or cosmic evolution, respectively, that are as yet undated but presumed to be connected via some hitherto unknown Markovian process (say, some yet-to-be-discovered form of stellar evolution).
Or at the opposite extreme, it might apply to processes on a very small scale (say, in particle or molecular physics) that leave a visible trace but are too fast or delicate to be tracked with a clock.
More exotically, such timeless data might be collected in a remote lab  without access to an external reference clock and with all internal clocks broken, so that the very notion of time has to be reconstructed from scratch.
In all these cases it is clearly impossible to construct the vector field $V$, and hence to infer the tensor $T$, in the straightforward manner outlined above.

Nevertheless, with the help of assumptions to be discussed below, a reconstruction of the Markovian dynamics is feasible even under such adverse circumstances.
One important prerequisite is that the dynamics is genuinely irreversible, $\dot{S}>0$.
Then at least the temporal order of the snapshots is easily established; 
and the entropy being a strictly monotonic function of time, it may in a first step serve as a proxy for time.
With the help of this proxy one can construct a vector field $W$, related to the true $V$ by $V=\dot{S} W$ (Fig. \ref{figure1}(c)). 
The reconstruction task thus reduces to the problem of finding the local entropy production rate, $\dot{S}$.
In the following, this unknown entropy production rate shall be denoted by a separate letter, $\eta$.

The true, yet to be determined vector field $V$ must satisfy a consistency condition, Eq. (\ref{consistencycondition}).
Replacing $V$ by $\eta W$ in this condition and exploiting $dS(W)=1$ leads to a first constraint on $\eta$, 
\begin{equation}
	d\eta\cdot dS
	\approx
	-2\eta
	,
\end{equation}
where the scalar product $d\eta\cdot dS$ is short for $C^{-1}(d\eta,dS)$.
I presume that 
the non-dissipative part of the dynamics is governed by some effective Hamiltonian, and that 
at least in the vicinity of equilibrium, this effective Hamiltonian does not vary with the macrostate.
The macroscopic non-dissipative dynamics thus shares with the microscopic dynamics a common structure, namely, unitary evolution with a time-independent Hamiltonian.
Such structural invariance is not guaranteed \textit{a priori;} 
rather, it constitutes an extra assumption which, however, is justified for many systems.
Mathematically, by Eqs. (\ref{EffectiveHamiltonian}) and (\ref{TnablaV}), and taking into account the change of sign when going from $\partial^a$ to $d\lambda^a$ via the metric, Eq. (\ref{metricandduals}),  
this assumption means that there must exist an $H_{\rm eff}$ such that 
\begin{equation}
	[\nabla V]_-(dS,\cdot) \approx \langle i{\cal L}_{\rm eff}G_a\rangle_{\pi(\rho)} d\lambda^a
	.
\end{equation}
Here $[\nabla V]_-$ denotes the antisymmetrized gradient (``curl'') of $V$. 
Again replacing $V$ by $\eta W$, exploiting $dS(W)=1$, and using the previous constraint on $\eta$  leads then to the condition
\begin{equation}
	\textstyle\frac{1}{2} d\eta
	\approx
	\eta\omega - \langle i{\cal L}_{\rm eff}G_a\rangle_{\pi(\rho)} d\lambda^a
	.
\label{etacondition}
\end{equation}
On the right-hand side, the one-form
\begin{equation}
	\omega:= [\nabla W]_-(dS,\cdot) - W
\end{equation}
is determined by the observed proxy field $W$.
The effective Hamiltonian (up to an additive and multiplicative constant), and hence the effective Liouvillian and the expectation values $\langle i{\cal L}_{\rm eff}G_a\rangle$ (up to a multiplicative constant), can be inferred from the observed equilibrium state, $\sigma\propto\exp(-\beta_\sigma H_{\rm eff})$, or from the observed hyperplane $\Sigma$, respectively.

Mirroring the indeterminacy of the effective Liouvillian,
the entropy production rate $\eta$, too, is only determined up to a multiplicative constant.
When the Liouvillian is rescaled by a constant factor $c$, ${\cal L}_{\rm eff}\to c {\cal L}_{\rm eff}$, then so is the entropy production rate, $\eta\to c \eta$.
Up to this undetermined multiplicative constant, the above condition, Eq. (\ref{etacondition}), specifies $\eta$ uniquely.
This enables one to convert the proxy field $W$ to the true vector field $V$.
From there, the reconstruction of the dynamics can proceed as before.


\section{Example}
\label{example}

In this section, I apply the general framework to a simple example.
In fact, the example is so simple that its dynamics might just as well be analyzed without the whole apparatus of differential geometry introduced above.
However, it serves to illustrate the internal consistency of the approach, and has the advantage of being solvable analytically.

To be specific, I consider an exchangeable assembly of qubits, possibly interacting with each other and with some unknown environment. 
(A physical realization might be a para- or ferromagnet.) 
The three Pauli operators $\sigma_x$, $\sigma_y$, and $\sigma_z$ constitute the relevant observables;
the associated manifold of macrostates is thus the Bloch sphere.
Measurements on samples taken from the assembly yield the expectation values of the relevant observables,
$x:=\langle\sigma_x\rangle$, $y:=\langle\sigma_y\rangle$, and $z:=\langle\sigma_z\rangle$.
Driven by some Markovian dynamics whose form is yet to be inferred, these expectation values change, tracing out orbits in the Bloch sphere.
The geometric shape of these orbits is given but ---for lack of time information--- their parametrization is not.

\begin{figure}[htbp]
\begin{center}
\includegraphics[width=8cm]{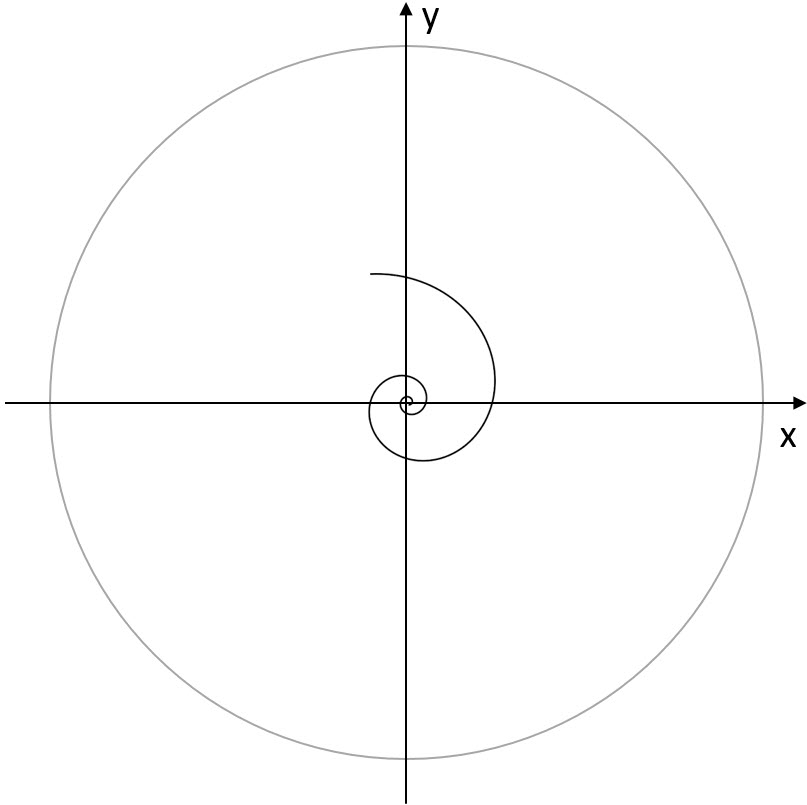}
\end{center}
\caption{Two-dimensional section ($z=0$) of the Bloch sphere.
Near equilibrium, $\sigma=I/2$, orbits have the form of a logarithmic spiral.
}
\label{spiral}
\end{figure}

For the sake of concreteness, I presume that whenever $z=0$ initially, it is $z=0$ on the entire orbit;
so orbits with this initial condition are constrained to the hyperplane $\Sigma=\{\mu\in\pi({\cal S})|z=0\}$.
Within this hyperplane, orbits are spiralling towards the equilibrium state, $\sigma$, which I take to be the totally mixed state, $\sigma=I/2$ (Fig. \ref{spiral}).
I assume that near equilibrium the orbits have the shape of a logarithmic spiral,
\begin{equation}
	\phi=\phi_0 + \ln r
	,
\end{equation}
where $r,\phi$ are polar coordinates in $\Sigma$,
\begin{equation}
	x=r\cos\phi
	\ ,\ 
	y=r\sin\phi
	.
\end{equation}
Given only this information, is it possible to reconstruct the dynamical law near equilibrium?

Near equilibrium the entropy function is approximately quadratic,
\begin{equation}
	S\approx \ln 2 - r^2 / 2
	.
\end{equation}
With entropy as a proxy for time, the orbit can be given a parametric representation,
\begin{equation}
	r(S)=\sqrt{2\ln 2-2S}
	\ ,\ 
	\phi(S)=\phi_0 + \textstyle\frac{1}{2}\ln(2\ln 2-2S)
	,
\end{equation}
which gives rise to the proxy field
\begin{equation}
	W= - \frac{1}{r}\frac{\partial}{\partial r} - \frac{1}{r^2}\frac{\partial}{\partial\phi}
	.
\end{equation}
Mapping vectors to one-forms with the help of the metric tensor, given near equilibrium by
\begin{equation}
	C\approx dr\otimes dr + r^2 d\phi\otimes d\phi
	,
\end{equation}
yields the one-form
\begin{equation}
	\omega=\frac{1}{r}dr+d\phi
	.
\end{equation}
From the orientation of the hyperplane $\Sigma$ one readily concludes that $H_{\rm eff}\propto\sigma_z$.
Exploiting moreover that near equilibrium it is $\lambda^x\approx -x$ (and likewise for $\lambda^y$), one finds
\begin{equation}
	\langle i{\cal L}_{\rm eff}G_a\rangle_{\pi(\rho)} d\lambda^a
	\approx
	c\,r^2 d\phi
	,
\end{equation}
where $c$ is some undetermined multiplicative constant.
Inserting these results into the equation for the entropy production rate, Eq. (\ref{etacondition}), yields
\begin{equation}
	\frac{1}{2}d\eta \approx \frac{\eta}{r}dr + (\eta-c\,r^2)d\phi
	,
\end{equation}
which has the unique solution
\begin{equation}
	\eta=c\,r^2
	.
\end{equation}
This entropy production rate allows one to convert the proxy field $W$ to the ``true'' vector field $V$,
\begin{equation}
	V=-c \left[r\frac{\partial}{\partial r} + \frac{\partial}{\partial\phi}\right]
	.
\end{equation}
Finally, by Eq. (\ref{TnablaV}), this yields the tensor field $T$ governing the Markovian dynamics, with the antisymmetric, non-dissipative part
\begin{equation}
	T^{(-)}\approx \frac{c}{r}\left[ \frac{\partial}{\partial r}\otimes \frac{\partial}{\partial\phi} - \frac{\partial}{\partial\phi}\otimes \frac{\partial}{\partial r} \right]
\end{equation}
and the symmetric, dissipative part
\begin{equation}
	T^{(+)}\approx c\,C^{-1}
	.
\end{equation}


\section{Discussion}
\label{discussion}

In the preceding sections I showed how it is possible, in principle, to infer a Markovian equation of motion --- with both its non-dissipative and dissipative parts ---  from experimental data even when the latter lack information about time.
To achieve this inference, I exploited the Riemannian geometry of the manifold of macrostates, as well as the structure of generic Markovian dynamics.
In addition, I made a number of assumptions:
I presumed that
(i) energy is conserved;
(ii) the experimental data have been taken near equilibrium (limiting thus the validity of the inferred equation of motion to the vicinity of equilibrium, too);
(iii) the dynamics is genuinely irreversible, $\dot{S}>0$;
(iv) the equilibrium is canonical; 
and 
(v) the effective, time-independent Hamiltonian featuring in this canonical equilibrium state also governs the effective non-dissipative dynamics near equilibrium.
The last assumption allows one to infer immediately the non-dissipative part of the Markovian dynamics (up to a multiplicative constant);
one simply identifies the effective Hamiltonian with the logarithm of the equilibrium state or, equivalently, with the observable that is conserved in the hyperplane $\Sigma$.
There remains then the task of inferring the dissipative part of the dynamics.
This requires more effort but is feasible, too, with the tools provided above.

The reconstruction scheme fixes time intervals only up to a multiplicative constant, and hence time itself only up to an affine transformation.
This freedom ensures the compatibility of the reconstruction scheme with special relativity.
Two observers, moving relative to each other, may see the same unparametrized orbits traced out by a Markovian process in the manifold of macrostates.
In the example discussed in Sec. \ref{example}, one observer might be in the rest frame of the experiment, while the other observer moves relative to this frame in the $z$ direction;
both observers see the same logarithmic spiral depicted in Fig. \ref{spiral}.
Their respective reconstructions of time are allowed to differ by an affine transformation.
This provides, in particular, for the possibility that they differ by a Lorentz transformation, as demanded by relativity.

The time reconstructed via the above scheme is a {\textit{macroscopic}} time,
for it is inferred from experimental data on some coarse-grained level of description.
If the same process is observed on two different levels of description, and if on both levels the process is Markovian, then the above procedure can be applied to both sets of data, yielding respective macroscopic times.
One may envision such a situation when one observes, say, a fluid on either the Boltzmann or the Navier-Stokes level of description; 
or when one observes a spin system on different coarse-grained length scales, corresponding to varying block spin sizes.
As long as the assumptions spelt out above hold on all these levels of description, the various macroscopic times agree (up to affine transformations).
The inferred time is thus invariant under a change of macroscopic scale, provided that on all scales the dynamics falls within the basic structure of generic Markovian dynamics;
and provided that, moreover, on all scales there exists an effective Hamiltonian for its non-dissipative part.
In this sense, the scale invariance of macroscopic time presupposes renormalizability \cite{PhysRevE.79.021124}.

The results in the present paper show that, in principle, any Markovian process can serve as a clock.
Absent an external reference clock, one may take an arbitrary, genuinely irreversible Markovian process, apply the above reconstruction scheme, and with the help of the inferred equation of motion, mark points on the orbits such that a segment between two successive points corresponds to some fixed time interval.
When the Markovian process is repeated, the system evolves along one of the orbits;
and whenever its macrostate reaches one of the designated points, the clock ``ticks.''
In contrast to everyday clocks, which are based on reversible, periodic motion,  such a clock would be based on the irreversibility of the process involved --- similar to, e.g., radiocarbon dating.

Conceptually, one may wonder why a reconstruction of time is necessary at all, and why one does not simply stick to entropy as a proxy for time.
In fact, any strictly monotonic function of time --- like entropy in a genuinely irreversible process --- serves the purpose of ordering events; 
and if such an alternative parameter were adopted universally, it would still be possible to keep appointments.
What distinguishes physical time from all other conceivable parameters is that when expressed as a function of time, equations of motion become particularly {simple}.
As Henri Poincar\'e put it succinctly \cite{poincare:time},
\begin{quote}
Time should be so defined that the equations of mechanics may be as simple as possible.
In other words, there is not one way of measuring time more true than another;
that which is generally adopted is only more \textit{convenient}.
\end{quote}
Determining physical time is thus tantamount to finding the simplest possible parametrization for the widest possible range of processes;
where ``simplest'' and ``widest'' must be suitably operationalized.
On a speculative note, the reconstruction scheme expounded here may be viewed as a proposal for such an operationalization:
``Simplicity'' would be embodied in the assumptions summarized in the opening paragraph of this discussion;
whereas ``widest'' would signify the scale invariance alluded to above.
Physical time would then be singled out as being a fixed point under changes of scale.

I see several avenues for further research.
First, it will be important to apply the reconstruction scheme to real or simulated experimental data that pertain to more complex, real-world problems.
Secondly, the framework should be extended to deal with processes further away from equilibrium.
And finally, it might be worthwhile to explore in more detail the conceptual issues raised in this discussion, i.e., the relativistic covariance of the procedure, its invariance under a change of scale, and the possible import of these findings on the fundamental definition and meaning of time.

\begin{acknowledgments}
I thank 
Hans Christian {\"O}ttinger 
for 
helpful discussions about the two-generator formulation of Markovian nonequilibrium dynamics.
This work was supported by the EU Integrating Project SIQS, the EU STREP EQUAM, and the ERC Synergy grant BioQ.
\end{acknowledgments}



%

\end{document}